# A novel methodology to assess the scientific standing of nations at field level[1]

*Giovanni Abramo*[*]
  Laboratory for Studies of Research and Technology Transfer
  Institute for System Analysis and Computer Science (IASI-CNR)
  National Research Council of Italy
  Viale Manzoni 30, 00185 Rome - ITALY
  giovanni.abramo@uniroma2.it

*Ciriaco Andrea D'Angelo*
  Department of Engineering and Management
  University of Rome "Tor Vergata"
  Via del Politecnico 1, 00133 Rome - ITALY
  dangelo@dii.uniroma2.it

**Abstract**

The formulation of national research policies would benefit greatly from reliable strategic analysis of the scientific infrastructure, aimed at identifying the relevant strengths and weaknesses at field level. Bibliometric methodologies thus far proposed in the literature are not completely satisfactory. This work proposes a novel "output-to-input-oriented" approach, which permits identification of research strengths and weaknesses on the basis of the ratios of top scientists and highly cited articles to research expenditures in each field. The proposed approach is applied to the Italian academic system. 2012-2016 scientific publications are analyzed, in the 218 research fields where bibliometric assessment is appropriate.

# 1. Introduction

An increasing number of countries place priority on strengthening their scientific infrastructure, in support of competitiveness and social development. However, with the limitations on public resources and increasing social needs, governments are also more attentive to the efficiency and effectiveness of their interventions. Ideally, policy-makers would seek effectiveness through identification and support of the scientific fields promising the highest social returns.

The first step then is spotting the scientific fields where the country's research outstands and those where it lags behind. The second step is assessing the alignment between the standing of a research field and its strategic relevance, in terms of the long-term economic competitiveness of the country and social development. This process will inform the formulation of research policies and action plans, and according resource allocation. In the case weak fields result as strategic, initiatives and funds should be devoted to turn them into strong fields.

There are a number of studies on the scientific standing of research systems. Unsatisfied with the analytical methods and indicators for this task thus far proposed in the literature, in 2014 we presented a new methodology to assess national strengths and weaknesses in research fields (Abramo, & D'Angelo, 2014). In this work, we revisit that method and indicators in the light of the advancements occurring in scientometrics and our further inquiries into the issue. The main results and relevant methodology of this study have been included in the 2019 Report on Research and Innovation in Italy, published by the National Research Council of Italy, and presented to the Prime Minister and the Minister of Education, Universities, and Research in a public event on 15 Oct. 2019.[2]

In the next section, we provide a review of the literature on the topic. In Section 3, we report the rationale and main features of the methodology that we proposed in 2014, and present the revised methodology. Sections 4 and 5 present the findings, respectively at the discipline and field level. The last section presents our conclusions and recommendations.

## 2. A review of the literature on measuring the scientific standing of nations.

Measuring the "scientific standing" of nations at field level is a complex task (Hauser and Zettelmeyer, 1997; Werner and Souder, 1997). Opinions on what "scientific standing" means are not unanimous, and less so the appropriate way to measure it. For sure, scientific standing implies a comparison, "surpassing something or someone in quality" (Tijssen, 2003).

The question then is what research quality is and if it differs from research impact. According to Boaz and Ashby (2003), and OECD (1997) impact is only one dimension of research quality, the other dimensions being relevance and rigor of research (Martin and Irvine, 1983). Others affirm that quality and impact are distinct elements of scientific standing (Grant et al., 2010), and still others that quality and impact are technically synonymous (Abramo, 2018).

The chronological analysis of studies measuring the scientific standing of nations reflect the developments occurred in bibliometric indicators and methods in the recent years. The milestone study by May (1997) compared the relative scientific standing of 15 nations in STEM, by the shares of WoS-indexed publications, citations, and citations per unit of spending, over a 14-year period. A year later, Adams (1998) compared England performance in 47 fields with that of six other countries. An update and extension of May's work (1997) was conducted by King (2004) who analyzed 31 countries over a 10-year period. The new study increased the number of nations analyzed (31), provided a longitudinal analysis over two five-year periods. King (2004) adopted additional bibliometric indicators, i.e. top 1% highly cited articles, HCAs, and mean citations per paper, and most importantly, used field-normalized citations. To account for the different sizes of countries, he also divided the performance indicators by the total number of researchers, R&D expenditures, and GDP.

---
[2] http://www.dsu.cnr.it/relazione-ricerca-innovazione-2019/index.html, last access on 8 November 2019.

More recently, works on the assessment of the relative research standing of nations have increasingly focused on excellence, as measured by HCAs. HCAs "provide a useful analytical framework – both in terms of transparency, cognitive and institutional differentiation, as well as its scope for domestic and international comparisons" (Tijssen, Visser, & Van Leeuwen, 2002). Pislyakov and Shukshina (2012) recurred to HCAs to identify "centers of excellence" in Russia, and to co-authorship analysis and investigate their collaboration with each other and with institutions abroad. Bornmann and Leydesdorff (2011) conceived a new mapping approach using Google Maps to identify which cities produce more excellent papers than can be expected. Bornmann, Leydesdorff, Walch-Solimena, & Ettl (2011) applied the same methodology for mapping field-specific centers of excellence around the world. Finally for some years, Nature has been publishing the Nature Index Annual Tables[3] highlighting the institutions and countries which dominated research across all sectors. Also such scientometric research groups as the CWTS of the University of Leiden[4], or the SCImago group[5] yearly publish on-line country rankings, using total number of publications, mean field-normalized citations per article, and share of HCAs out of total output.

Our greatest concern over the approaches proposed in the literature is that either they do not apply efficiency (output-to-input) indicators at all, or when they try to do it, they fail to account for the different intensity of publications, and use of inputs across research fields. Most studies in fact fail to establish a relation between the inputs that a country employs in each field and the relevant output and impact. In all the literature that adopts only output indicators (total number or world share of publications, citations, and HCAs), it is no surprise that the USA invariably ranks at the top, in all scientific fields.

Those studies do not answer the question whether the USA is at the top because it invests in research more than other countries, or because scientists in USA perform better than colleagues abroad. The attempts to answer the question by normalizing outputs and impact to input, have used overall input data at the aggregate level, ignoring the sectoral differences of the knowledge production function (Abramo and D'Angelo, 2007). Apart from rare exceptions, bibliometricians in fact generally lack all or part of data on the identity of researchers, their field of research, and the research expenditures per field in the individual countries under comparison.

The attempt to circumvent the lack of input data with size independent indicators of the kind "average citations per publication", results ineffective, as such indicators violate an axiom of production theory: as output increases under equal inputs, performance cannot be considered to diminish (Abramo & D'Angelo, 2016a; Abramo & D'Angelo, 2016b). With the "average citations per paper" approach, this violation occurs when any organization (or individual) produces an additional publication under equal inputs, whose (normalized) impact is even slightly below their previous average.

---

[3] https://www.natureindex.com/news-blog/top-ten-countries-research-science-twenty-nineteen, last accessed on 8 November 2019
[4] http://www.leidenranking.com/ranking, last accessed on 8 November 2019.
[5] http://www.scimagojr.com/countryrank.php, last accessed on 8 November 2019.

To the best of our knowledge, apart from Italy, the only successful attempts so far to account for inputs when assessing individual, field, and institutional performance at national level, are the ones concerning Norway (Abramo, Aksnes, & D'Angelo, 2019) and Sweden (Sandström, & Van den Besselaar, 2018). While the former deploys an approach based on the actual levels of inputs and outputs, the latter relies on the change in the levels of inputs and outputs: an approach that to a large extent eliminates the problem of measurement differences between countries. Informally, we have also come to know that Poland will soon be ready to follow suit.

## 3. A new methodology

Our methodology is based on the assertion that at country level a research field is stronger than another one if researchers in that field perform better than the ones in the other field. Because the intensity of publication is not homogenous across fields (D'Angelo & Abramo, 2015), the direct comparison of performance between fields would favour fields with higher intensity of publication, with noticeable distortions of results (Abramo, Cicero, & D'Angelo, 2013). The correct way to proceed then would be to compare the performance of a country's researchers in a field with those of the same field in other countries, and then compare positions in the rankings of different fields.

The ideal indicator for measuring and comparing performance across fields is research productivity[6] at the field level, but to calculate such an indicator requires knowledge of the field of researcher of each scientist, as well as their output (Abramo & D'Angelo, 2014). Unfortunately, the data on scientists in the different nations (where available) in general do not include their fine-grained classification by field of research, with the sole exception of Italy, to the best of our knowledge.

Since we cannot measure the research productivity at field level of each nation, in a previous work of ours (Abramo & D'Angelo, 2014) we devised an alternative method based on the classification of individual researchers by scientific field, and their highly cited articles (HCAs). The methodology is described below.

Our methodology took advantage of a characteristic that, as we said, seems unique to the Italian research system, in which each academic is officially classified as belonging to a single specifically-defined research field, called "Scientific Disciplinary Sector" (SDS). The national academic system is composed of 370 SDSs, grouped into 14 "University Disciplinary Areas" (UDAs). In other countries lacking a similar system it would still be possible to attempt to classify researchers by disambiguating their output, and identifying the prevalent subject category of their publications. With such a classification for a range of countries, it would be possible to compare a proxy of productivity in each field over specific periods. The recent development of research information management systems (RIMs) and their adoption by an increasing number of countries, is a ray of hope for a near future availability of similar datasets in other countries (Sandstrom, Sandstrom, & Van den Besselaar, 2019; Kulczycki, ..., & Zuccala, 2018). Because such classifications and datasets have not yet been developed in the largest countries, for a robust comparison, we devised a different route to answer our research question.

---

[6] In simple terms, we define research productivity in a given period as total impact divided by the cost of input.

We resorted to the HCAs, which we defined as those publications that place in the top 5% or 10% (HCAs$_{(5\%)}$ or HCAs$_{(10\%)}$) of the world citation rankings for Web of Science (WoS) indexed publications[7] of the same year and subject category.[8] For fields of comparable intensity of publication, we qualified one field as stronger than another if the ratio of researchers publishing HCAs to total in the field was higher. The underlying rationale was that the higher the concentration of researchers in a field who can produce HCAs, meaning capable of notably advance the frontier of knowledge in that field, the relatively stronger is that field in the country.

Our methodology embedded a severe limit though, as comparisons were possible only for fields of comparable intensity of publication. Without that condition, fields with higher publication intensity would be favored, because the probability of having an article among the highly cited ones increases with the number of articles produced, which depends on both the quality of a scientist and the average intensity of publication in the field.

The novel methodology that we propose in this work overcomes this limit by normalizing HCAs in a field by the intensity of publication in that field by top scientists (TSs), defined as researchers with a total fractional counting of HCAs (referred to as FHCAs when the fractional counting method is applied) that exceeds a certain threshold. We also adopt fractional counting to control for the intensity of joint research work, which varies across fields (Abramo, D'Angelo, & Murgia, 2013). Furthermore, instead of measuring the ratio of TSs in a field to the total number of researchers in that field, here we measure the number of TSs per euro spent in research in the field. Finally, we add one more indicator, i.e. the average number of FHCAs (normalized by the intensity of publications of TSs) per euro spent in research in the field. The first indicator measures the spread of the capability to noticeably advance the frontiers of knowledge among the scientists in a field; the second measures the extent of such outstanding contribution.

In the following we apply the above two indicators to the assessment of the research strengths and weaknesses of the Italian academic system. The observation period is 2012-2016 and citations are counted on 30 October 2018.

We extract data on the Italian faculty from the database on university personnel, maintained by the Italian Ministry of Education, Universities, and Research. For each professor,[9] this database provides information on their name and surname, gender, affiliation, field classification and academic rank, at close of each year.[10]

Drawing on the classification of all Italian professors in their research fields, from the raw data of the WoS publications over the period 2012-2016, and applying an algorithm for reconciliation of the author's affiliation and disambiguation of their precise identity, we attribute each publication to the university researcher that produced it[11] (D'Angelo, Giuffrida, & Abramo, 2011). We then identify all HCAs, and divide each of them by the number of authors, to get the FHCA value for each HCA.

---

[7] For publications in multi-category journals, we considered the percentile for the most favourable category.
[8] It would also be possible to choose thresholds other than 1% and 5%. Glänzel and Schubert (1988) provide further discussion.
[9] Unfortunately, we cannot include scientists from research institutions, because they lack SDS classification.
[10] http://cercauniversita.cineca.it/php5/docenti/cerca.php, last accessed on 8 November 2019.

As for research expenditure, production factors (input) consist of labor and capital (all resources other than labor, such as scientific instruments, databases, buildings, etc.). Information on individual salaries is unavailable in Italy but the salary ranges for ranks and seniority are published.[12] We are then able to approximate the salary for each individual as the national average of their academic rank.

The cost of capital per R&D man year is not available in Italy but it is in Norway, where it averages 42,693 euro PPP.[13] We assume then that Italian professors can count on the same amount of resources to conduct research. The further assumption is that capital is equally available to each professor, regardless of academic rank, research field,[14] and university.

We denote as $FSS_{TS}$ the scientific strength (F stands for the fractional counting method) of a field, measured by the number of TSs per cost of research, in formulae:

$$FSS_{TS} = \frac{1}{t} \cdot \frac{TSs}{\sum_{i=1}^{N}\left(\frac{w_r}{2} + k\right)}$$

[1]

where:
$TSs$ = number of professors with an outstanding FHCA score (i.e. above $Q_3 + 1.5 \times IQR$)[15] in the SDS, in the period under observation;
$w_r$ = average yearly salary of professor of academic rank $r$;
$k$ = yearly capital available for research to each professor, regardless of academic rank;
$t$ = number of years of work by the professor in the period under observation;
$N$ = number of professors in the field in the period under observation.

We halve labor cost, because we assume that 50% of professors' time is allocated to activities other than research (teaching, technology transfer, administration).

Table 1 summarizes cost of labor, cost of capital and total cost normalization factor per academic rank. In the following, we will use the total cost normalization factor to report measures of productivity.

---

[11] The harmonic mean of precision and recall (F-measure) of authorships disambiguated by our algorithm is around 97% (2% margin of error, 98% confidence interval).
[12] CINECA-Dalia, https://dalia.cineca.it/php4/inizio_access_cnvsu.php, last accessed on 8 November 2019
[13] Source: The R&D Statistics Bank, NIFU:
http://www.foustatistikkbanken.no/nifu/index.jsp?submode=default&mode=documentation&top=yes&language=en. Last accessed on 8 November 2019.
[14] It is well known that certain research fields require much more capital than others (e.g. mathematics vs physics), but it is correct to account for capital and make it equal to all fields, to avoid favouring less consuming fields, when comparing performance. The underlying assumption is that the capital made available to each researcher (regardless of its actual amount) is sufficient to carry out research in the relevant field.
[15] $Q_3$ is the value corresponding to the third quartile of the distribution, while IQR equals the difference between $Q_3$ and $Q_1$ values. This threshold is general used to identify true outliers in box plot for not normal distributions. Of course the threshold can be changed to suit different contexts. In ours, we found that it helps reduce drastically the correlation between $FSS_{TS}$ and the intensity of fractional publications in each field.

*Table 1: Production factor costs (euro) by academic rank*

| Academic rank | $w_r$ | $k$ | $\frac{w_r}{2} + k$ | Total cost normalization factor |
|---|---|---|---|---|
| Assistant professors | 54628 | 42693 | 70007 | 1 |
| Associate professors | 66821 | 42693 | 76104 | 1.09 |
| Full professors | 101301 | 42693 | 93344 | 1.23 |

Similarly, we denote as $FSS_{FHCA}$ the scientific strength of a field, measured by the number of FHCAs per cost of research, in formulae:

$$FSS_{FHCA} = \frac{1}{t} \cdot \frac{FHCAs^R}{\sum_{i=1}^{N}\left(\frac{w_r}{2} + k\right)}$$

[2]

where:
$FHCAs^R$ = total number of FHCAs published by professors in the field in the period under observation, rescaled by the average fractional publications by TSs, in the field.

$FSS_{TS}$ denotes the spread of the capability to noticeably advance the frontier of knowledge among the professors in a field, per euro spent in research; while $FSS_{FHCA}$ denotes the size of the overall contribution to noticeable advancement of the frontier of knowledge.

Both of these fine-grained field-level indicators can then be aggregated at discipline level, weighting each field by its size in terms of total R&D expenditures.

For reasons of robustness, this study is limited to those fields where bibliometric analysis can be considered significant, i.e. the sciences and social sciences. We exclude arts and humanities, where WoS coverage of publications is too limited to assure robust results.

We thus analyze 218 SDSs, belonging to 11 UDAs. The 218 SDSs include roughly 39,000 professors on faculty for at least three years over the 2012-2016 period,[16] who produced more than 300,000 WoS-listed publications.

Table 2 presents the distribution of SDSs, professors and publications per UDA. The data show the predominance of Medicine concerning all the dimensions reported. Researchers in this discipline alone represent 24.4% of the total dataset, producing 31.5% of the publications, with 34.8% of total HCAs$_{(5\%)}$ and 33.1% of HCAs$_{(10\%)}$. Both HCAs$_{(5\%)}$ and HCAs$_{(10\%)}$ are greater than expected.

Concerning the HCAs$_{(5\%)}$, we observe that they represent 8.0% of total publications in the dataset, with peaks in Physics (10.8%), Pedagogy and psychology (9.1%), Medicine and Industrial and information engineering (both with 8.8%). Considering instead the HCAs$_{(10\%)}$ (overall representing 15.5% of total publications), the UDA with highest incidence remains Physics (19.6%), followed by Pedagogy and psychology (18.2%), and Chemistry (17.2%).

---

[16] See Abramo, D'Angelo, and Cicero (2012) for details about this choice.

*Table 2: Dataset for the analysis, per discipline, UDA (data 2012-2016)*

| UDA | No. of SDSs | No. of professors | Publications | HCAs(5%) | HCAs(10%) |
|---|---|---|---|---|---|
| Mathematics and computer sciences | 10 | 3125 | 21090 | 1148 (5.4%) | 2132 (10.1%) |
| Physics | 8 | 2215 | 27232 | 2928 (10.8%) | 5350 (19.6%) |
| Chemistry | 11 | 2887 | 29678 | 2274 (7.7%) | 5093 (17.2%) |
| Earth sciences | 12 | 1042 | 8300 | 464 (5.6%) | 1070 (12.9%) |
| Biology | 19 | 4803 | 39854 | 3246 (8.1%) | 6699 (16.8%) |
| Medicine | 50 | 9637 | 94491 | 8332 (8.8%) | 15403 (16.3%) |
| Agricultural and veterinary sciences | 30 | 2998 | 19690 | 972 (4.9%) | 2257 (11.5%) |
| Civil engineering | 9 | 1510 | 11959 | 908 (7.6%) | 1866 (15.6%) |
| Industrial and information engineering | 42 | 5245 | 65781 | 5774 (8.8%) | 10804 (16.4%) |
| Pedagogy and psychology | 10 | 1410 | 9687 | 884 (9.1%) | 1765 (18.2%) |
| Economics and statistics | 17 | 4549 | 13302 | 841 (6.3%) | 1573 (11.8%) |
| Total | 218 | 39421 | 300274† | 23917† (8.0%) | 46468† (15.5%) |

*† The total value is different than the sum of values per column due to multiple counting of publications co-authored by Italian professors in different UDAs.*

## 4. Results

In this section we present the results of the analysis for both indicators, at discipline level first and field level then.

### 4.1 Strengths and weaknesses at discipline level

Concerning the indicator defined in [1], the weighted aggregation of scores for the SDSs of each UDA provides the values shown in Table 3. At the overall level, the average value of $FSS_{TS(5\%)}$ is 5.53. There are 4 UDAs with performance superior to that benchmark, led by Chemistry (7.20) and followed by Medicine (7.14), Industrial and information engineering (7.09) and Biology (6.80). Relaxing the threshold for HCAs to the top 10%, the ordering varies slightly, but Chemistry (9.21) continues its outstanding performance, tied with Biology, with these two followed by Earth sciences (8.65) and Medicine (8.45).

According to the indicator defined in [2], the relative performance of the disciplines varies slightly, as seen in Table 4. Independent of the cutoff for HCAs, Physics still leads (13.42 and 25.96), with Chemistry in second place for $FSS_{FHCA(10\%)}$(23.42) and Industrial and information engineering for $FSS_{FHCA(5\%)}$(11.04). In the lower part of the ranking (last five positions) the ordering remains unchanged with the two cutoffs of HCAs.

*Table 3: Number of top scientists ($TSs_{(5\%)}$ and $TSs_{(10\%)}$) (percentages out of total professors in brackets) and Fractional Scientific Strength ($FSS_{TS(5\%)}$ and $FSS_{TS(10\%)}$) in each discipline (UDA)*

| UDA | $TSs_{(5\%)}$ | $TSs_{(10\%)}$ | $FSS_{TS(5\%)}$ | $FSS_{TS(10\%)}$ |
|---|---|---|---|---|
| Chemistry | 111 (3.8%) | 142 (4.9%) | 7.20 | 9.21 |
| Biology | 172 (3.6%) | 233 (4.9%) | 6.80 | 9.21 |
| Earth sciences | 22 (2.1%) | 48 (4.6%) | 3.97 | 8.65 |
| Medicine | 361 (3.7%) | 427 (4.4%) | 7.14 | 8.45 |
| Industrial and information engineering | 202 (3.9%) | 236 (4.5%) | 7.09 | 8.28 |
| Civil engineering | 34 (2.3%) | 64 (4.2%) | 4.17 | 7.86 |
| Pedagogy and psychology | 39 (2.8%) | 56 (4.0%) | 5.22 | 7.50 |
| Agricultural and veterinary sciences | 81 (2.7%) | 111 (3.7%) | 5.04 | 6.90 |
| Physics | 51 (2.3%) | 67 (3.0%) | 4.31 | 5.67 |
| Mathematics and computer sciences | 41 (1.3%) | 91 (2.9%) | 2.43 | 5.39 |
| Economics and statistics | 49 (1.1%) | 64 (1.4%) | 1.98 | 2.59 |
| Overall | 1,163 (3.0%) | 1,539 (3.9%) | 5.53 | 7.31 |

*Table 4: Fractional Scientific Strength ($FSS_{FHCAs(5\%)}$ and $FSS_{FHCAs(10\%)}$) in each discipline (UDA)*

| UDA | $FSS_{FHCAs(5\%)}$ | $FSS_{FHCAs(10\%)}$ |
|---|---|---|
| Physics | 13.42 | 25.96 |
| Chemistry | 8.86 | 23.42 |
| Industrial and information engineering | 11.04 | 22.63 |
| Pedagogy and psychology | 9.36 | 21.59 |
| Biology | 8.14 | 20.79 |
| Civil engineering | 8.35 | 19.35 |
| Medicine | 7.39 | 16.99 |
| Earth sciences | 4.91 | 14.69 |
| Agricultural and veterinary sciences | 4.50 | 12.27 |
| Mathematics and computer science | 4.14 | 9.09 |
| Economics and statistics | 4.03 | 8.72 |
| Overall | 7.59 | 17.41 |

## 4.2 Strengths and weaknesses at field level

For a strategic analysis to appropriately serve research policy, a finer-grained assessment at field level is required. In this subsection, we present an extract of the results concerning strong and weak fields within and across disciplines. Obviously the fields can have very different sizes: as an illustrative example, in Table 5 we show the number of professors (total and divided in three academic ranks) and the total cost[17] for the 20 largest and 20 smallest SDSs.

---

[17] We recall that the total cost is a function of the size of the research staff and academic ranking, but not of capital, which we have assumed to be equal to each scientist.

*Table 5: Research staff and total cost for the 20 largest and 20 smallest scientific disciplinary sectors (SDSs)*

| SDS | UDA* | Assistant prof. | Associate prof. | Full prof. | Total | Total cost (× 100k euro) |
|---|---|---|---|---|---|---|
| INF/01 - Computer Science | 1 | 263 | 336 | 247 | 846 | 3180 |
| BIO/10 - Biochemistry | 5 | 293 | 304 | 230 | 827 | 3077 |
| FIS/01 - Experimental Physics | 2 | 175 | 393 | 239 | 807 | 3009 |
| MED/18 - General Surgery | 6 | 346 | 287 | 181 | 814 | 2975 |
| MAT/05 - Mathematical Analysis | 1 | 184 | 315 | 282 | 781 | 2965 |
| MED/09 - Internal Medicine | 6 | 317 | 296 | 188 | 801 | 2934 |
| SECS-P/01 - Political Economy | 13 | 167 | 255 | 330 | 752 | 2898 |
| SECS-P/07 - Business Administration | 13 | 193 | 300 | 258 | 751 | 2848 |
| ING-INF/05 - Data Processing Systems | 9 | 180 | 303 | 230 | 713 | 2702 |
| BIO/14 - Pharmacology | 5 | 243 | 226 | 180 | 649 | 2408 |
| CHIM/06 - Organic chemistry | 3 | 181 | 221 | 170 | 572 | 2147 |
| SECS-P/08 - Corporate Finance | 13 | 155 | 205 | 203 | 563 | 2133 |
| BIO/09 - Physiology | 5 | 212 | 208 | 154 | 574 | 2127 |
| CHIM/03 - General and Inorganic Chemistry | 3 | 135 | 242 | 151 | 528 | 1973 |
| MED/04 - General Pathology | 6 | 193 | 161 | 155 | 509 | 1886 |
| CHIM/08 - Pharmaceutical Chemistry | 3 | 140 | 207 | 90 | 437 | 1623 |
| SECS-S/06 - Mathematics for economics, act. studies and finance | 13 | 120 | 157 | 153 | 430 | 1622 |
| FIS/03 - Physics of matter | 2 | 101 | 193 | 137 | 431 | 1614 |
| MED/28 - Odonto-Stomalogical Diseases | 6 | 146 | 164 | 115 | 425 | 1577 |
| SECS-S/01 - Statistics | 13 | 95 | 176 | 139 | 410 | 1552 |
| … | | | | | | |
| MED/47 - Nursing and Midwifery | 6 | 4 | 2 | 0 | 6 | 21 |
| ING-IND/30 - Hydrocarburants and Fluids of the Subsoil | 9 | 2 | 5 | 1 | 8 | 30 |
| ING-IND/20 - Nuclear Measurement Tools | 9 | 2 | 4 | 4 | 10 | 39 |
| ING-IND/18 - Nuclear Reactor Physics | 9 | 2 | 7 | 3 | 12 | 47 |
| ING-IND/29 - Raw Materials Engineering | 9 | 4 | 9 | 1 | 14 | 52 |
| AGR/06 - Wood Technology and Woodland Management | 7 | 5 | 8 | 3 | 16 | 55 |
| ING-IND/02 - Naval and Marine construction and installation | 9 | 3 | 9 | 4 | 16 | 60 |
| GEO/12 - Oceanography and Atmospheric Physics | 4 | 5 | 7 | 7 | 19 | 73 |
| MED/48 - Neuropsychiatric and Rehabilitation Nursing | 6 | 11 | 7 | 5 | 23 | 83 |
| ING-IND/01 - Naval Architecture | 9 | 5 | 11 | 6 | 22 | 85 |
| ING-IND/23 - Applied Physical Chemistry | 9 | 3 | 7 | 12 | 22 | 86 |
| SECS-S/02 - Statistics for experimental and tech. research | 13 | 10 | 7 | 7 | 24 | 87 |
| ING-IND/07 - Aerospatial Propulsion | 9 | 5 | 11 | 9 | 25 | 93 |
| ING-IND/28 - Excavation Engineering and Safety | 9 | 10 | 9 | 7 | 26 | 96 |
| ING-IND/03 - Flight Mechanics | 9 | 4 | 12 | 11 | 27 | 104 |
| FIS/08 - Didactics and History of Physics | 2 | 6 | 18 | 6 | 30 | 109 |
| AGR/14 - Pedology | 7 | 9 | 13 | 8 | 30 | 111 |
| MED/45 - General, Clinical and Pediatric Nursing | 6 | 5 | 20 | 5 | 30 | 114 |
| ING-IND/05 - Aerospace Systems | 9 | 8 | 16 | 8 | 32 | 117 |
| MED/02 - History of Medicine | 6 | 17 | 8 | 9 | 34 | 123 |

\* *1, Mathematics and computer sciences; 2, Physics; 3, Chemistry; 4, Earth sciences; 5, Biology; 6, Medicine; 7, Agricultural and veterinary sciences; 8, Civil engineering; 9, Industrial and information engineering; 10, Pedagogy and psychology; 11, Economics and statistics*

Table 6 presents the example of the FSSs across all the SDSs in UDA Earth sciences. It can be observed the ample variability of FSS scores across fields within the same discipline.

*Table 6: Fractional Scientific Strength and size (total cost) of each field (SDS) falling in discipline (UDA) Earth sciences*

| SSD* | Total cost (x 100k euro) | $FSS_{TS}$ 5% | $FSS_{TS}$ 10% | $FSS_{FHCA}$ 5% | $FSS_{FHCA}$ 10% |
|---|---|---|---|---|---|
| GEO/01-Palaeontology and Palaeoecology | 364.8 | 1.92 | 1.92 | 2.34 | 8.77 |
| GEO/02-Stratigraphic and Sedimentological Geology | 519.4 | 2.70 | 5.39 | 4.52 | 13.67 |
| GEO/03-Structural geology | 394.2 | 0.00 | 14.21 | 5.78 | 17.60 |
| GEO/04-Physical Geography and Geomorphology | 460.6 | 7.60 | 7.60 | 8.53 | 22.31 |
| GEO/05-Applied geology | 469.0 | 8.96 | 13.44 | 7.00 | 17.33 |
| GEO/06-Mineralogy | 295.7 | 2.37 | 11.84 | 2.69 | 7.64 |
| GEO/07-Petrology and Petrography | 323.1 | 6.50 | 6.50 | 5.26 | 15.01 |
| GEO/08-Geochemistry and Volcanology | 285.8 | 4.90 | 9.80 | 4.08 | 19.56 |
| GEO/09-Mineral Geological Resources and Mineralogic and Petrographic Applications | 254.3 | 2.75 | 8.26 | 5.14 | 14.55 |
| GEO/10-Geophysics of Solid Earth | 261.9 | 0.00 | 10.69 | 1.75 | 8.40 |
| GEO/11-Applied geophysics | 181.7 | 0.00 | 0.00 | 2.14 | 6.96 |
| GEO/12-Oceanography and Atmospheric Physics | 72.9 | 9.60 | 19.19 | 7.54 | 21.23 |

Considering all 218 SDSs investigated, Table 7 presents the 10 strongest SDSs according to the indicator defined in [1], with 5% threshold, as well as the weakest, which all happen to have $FSS_{TS} = 0$ (36 in all). Table 8 instead presents the 10 strongest and 10 weakest SDSs according to the indicator defined in [2], again with 5% threshold.

The analysis by the indicator $FSS_{TS}$ reveals the peculiar presence of an SDS concerned with nuclear energy among the top 10, at more than 30 years since the Italian referendum that halted all nuclear power sites on national territory. The strongest ten SDSs belong to only three disciplines, i.e. Medicine, Agricultural and veterinary sciences; and Industrial and information engineering. A transverse reading of the $FSS_{FHCA}$ indicator discloses different facts, with Industrial and information engineering now represented by 5 SDSs, but with only one Medicine SDS (MED/15) among the strongest.

*Table 7: Strongest 10 and weakest fields (SDSs) by Fractional Scientific Strength ($FSS_{TS(5\%)}$)*

| SSD | UDA* | $FSS_{TS(5\%)}$ |
|---|---|---|
| AGR/08 - Agrarian Hydraulics and Hydraulic Forest Management | 7 | 19.01 |
| MED/48 - Neuropsychiatric and Rehabilitation Nursing | 6 | 16.90 |
| MED/15 - Blood Diseases | 6 | 15.50 |
| ING-IND/18 - Nuclear Reactor Physics | 9 | 15.02 |
| MED/10 - Respiratory Diseases | 6 | 14.86 |
| MED/11 - Cardiovascular Diseases | 6 | 14.58 |
| ING-IND/22 - Science and Technology of Materials | 9 | 14.53 |
| MED/03 - Medical Genetics | 6 | 14.48 |
| ING-IND/29 - Raw Materials Engineering | 9 | 13.49 |
| MED/49 - Applied Dietary Sciences | 6 | 13.03 |
| … | - | - |
| AGR/05 - Forestry and Silviculture | 7 | 0 |
| AGR/06 - Wood Technology and Woodland Management | 7 | 0 |
| AGR/10 - Rural Construction and Environmental Land Management | 7 | 0 |
| FIS/04 - Nuclear and Subnuclear Physics | 2 | 0 |
| FIS/06 - Physics for Earth and Atmospheric Sciences | 2 | 0 |
| FIS/08 - Didactics and History of Physics | 2 | 0 |
| GEO/03 - Structural Geology | 4 | 0 |
| GEO/10 - Geophysics of Solid Earth | 4 | 0 |
| GEO/11 - Applied Geophysics | 4 | 0 |
| ICAR/01 - Hydraulics | 8 | 0 |
| ING-IND/02 - Naval and Marine construction and installation | 9 | 0 |
| ING-IND/03 - Flight Mechanics | 9 | 0 |
| ING-IND/07 - Aerospatial Propulsion | 9 | 0 |
| ING-IND/12 - Mechanical and Thermal Measuring Systems | 9 | 0 |
| ING-IND/20 - Nuclear Measurement Tools | 9 | 0 |
| ING-IND/23 - Applied Physical Chemistry | 9 | 0 |
| ING-IND/26 - Theory of Development for Chemical Processes | 9 | 0 |
| ING-IND/28 - Excavation Engineering and Safety | 9 | 0 |
| ING-IND/30 - Hydrocarburants and Fluids of the Subsoil | 9 | 0 |
| MAT/01 - Mathematical Logic | 1 | 0 |
| MAT/04 - Complementary Mathematics | 1 | 0 |
| MAT/06 - Probability and Mathematical Statistics | 1 | 0 |
| MED/19 - Plastic Surgery | 6 | 0 |
| MED/20 - Pediatric and Infant Surgery | 6 | 0 |
| MED/27 - Neurosurgery | 6 | 0 |
| MED/31 - Otorinolaringology | 6 | 0 |
| MED/34 - Physical and Rehabilitation Medicine | 6 | 0 |
| MED/47 - Nursing and Midwifery | 6 | 0 |
| SECS-P/04 - History of Economic Thought | 13 | 0 |
| SECS-P/10 - Business Organisation | 13 | 0 |
| VET/03 - General Pathology and Veterinary Pathological Anatomy | 7 | 0 |
| VET/05 - Infectious Diseases of Domestic Animals | 7 | 0 |
| VET/07 - Veterinary Pharmacology and Toxicology | 7 | 0 |
| VET/08 - Clinical Veterinary Medicine | 7 | 0 |
| VET/09 - Clinical Veterinary Surgery | 7 | 0 |
| VET/10 - Clinical Veterinary Obstetrics and Gynaecology | 7 | 0 |

\* *1, Mathematics and computer sciences; 2, Physics; 3, Chemistry; 4, Earth sciences; 5, Biology; 6, Medicine; 7, Agricultural and veterinary sciences; 8, Civil engineering; 9, Industrial and information engineering; 10, Pedagogy and psychology; 11, Economics and statistics*

*Table 8: Strongest 10 and weakest 10 fields (SDSs) by Fractional Scientific Strength ($FSS_{FHCAs(5\%)}$)*

| SSD | UDA* | $FSS_{FHCAs(5\%)}$ |
|---|---|---|
| ING-IND/09 - Energy and Environmental Systems | 9 | 25.09 |
| MED/15 - Blood Diseases | 6 | 23.19 |
| ICAR/05 - Transport | 8 | 20.82 |
| FIS/05 - Astronomy and Astrophysics | 2 | 20.44 |
| ING-IND/11 - Environmental Technical Physics | 9 | 19.68 |
| ING-IND/27 - Industrial and Technological Chemistry | 9 | 18.82 |
| ING-IND/32 - Electrical Convertors, Machines and Switches | 9 | 16.63 |
| FIS/01 - Experimental Physics | 2 | 16.28 |
| ING-IND/25 - Chemical Plants | 9 | 16.28 |
| M-PSI/02 - Psychobiology and Physiological Psychology | 11 | 16.05 |
| … | - | - |
| ING-IND/30 - Hydrocarburants and Fluids of the Subsoil | 9 | 0.0 |
| MAT/04 - Complementary Mathematics | 1 | 0.0 |
| VET/08 - Clinical Veterinary Medicine | 7 | 0.18 |
| ING-IND/01 - Naval Architecture | 9 | 0.35 |
| VET/10 - Clinical Veterinary Obstetrics and Gynaecology | 7 | 0.43 |
| VET/09 - Clinical Veterinary Surgery | 7 | 0.54 |
| MAT/02 - Algebra | 1 | 0.64 |
| MED/20 - Pediatric and Infant Surgery | 6 | 0.71 |
| MED/29 - Maxillofacial Surgery | 6 | 0.80 |
| MED/31 - Otorinolaringology | 6 | 0.86 |

*\* 1, Mathematics and computer sciences; 2, Physics; 3, Chemistry; 4, Earth sciences; 5, Biology; 6, Medicine; 7, Agricultural and veterinary sciences; 8, Civil engineering; 9, Industrial and information engineering; 10, Pedagogy and psychology; 11, Economics and statistics*

Applying the two indicators of FSS and two variants for each ($HCAs_{(5\%)}$ and $HCAs_{(10\%)}$) we obtain four distributions. Table 9 presents the Spearman correlation values for the four distributions. For the indicator $FSS_{FHCA}$, the correlation between rankings emerging from the different thresholds of HCA (5% vs 10%) is very high (Spearman ρ equal to 0.960) and greater than that for the two thresholds of $FSS_{TS}$, which is still high (0.603). The correlation between the two indicators $FSS_{TS}$ and $FSS_{FHCA}$ is instead lower, with Spearman ρ equal to 0.500 for the 5% cutoff, and 0.428 for the 10% cutoff. This last observation reveals that in the large part of the SDSs there are relatively few TSs (outliers) who produce the large part of the FHCAs, or on the other hand many TSs, each producing relatively few HCAs.

*Table 9: Spearman correlation matrix of the four Fractional Scientific Strength indicators considered*

| | $FSS_{TS(5\%)}$ | $FSS_{TS(10\%)}$ | $FSS_{FHCA(5\%)}$ | $FSS_{FHCA(10\%)}$ |
|---|---|---|---|---|
| $FSS_{TS(5\%)}$ | 1 | 0.603 | 0.500 | 0.507 |
| $FSS_{TS(10\%)}$ | | 1 | 0.376 | 0.428 |
| $FSS_{FHCA(5\%)}$ | | | 1 | 0.960 |
| $FSS_{FHCA(10\%)}$ | | | | 1 |

In this regard, the following figures show the dispersion of the 218 SDSs considering the value of both the indicators with the cutoff for HCA at the top 5% (Figure 1) and top 10% (Figure 2) of articles.

*Figure 1: Dispersion of the two Fractional Scientific Strength indicators registered for the 218 SDSs analyzed, for top 5% cited articles*

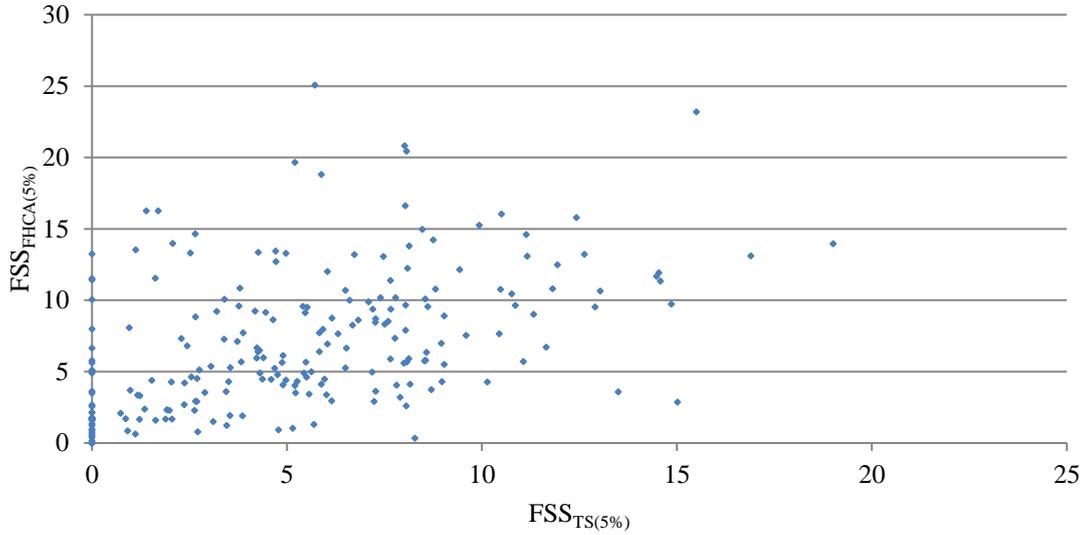

*Figure 2: Dispersion of the two Fractional Scientific Strength indicators registered for the 218 SDSs analyzed, for top 10% cited articles*

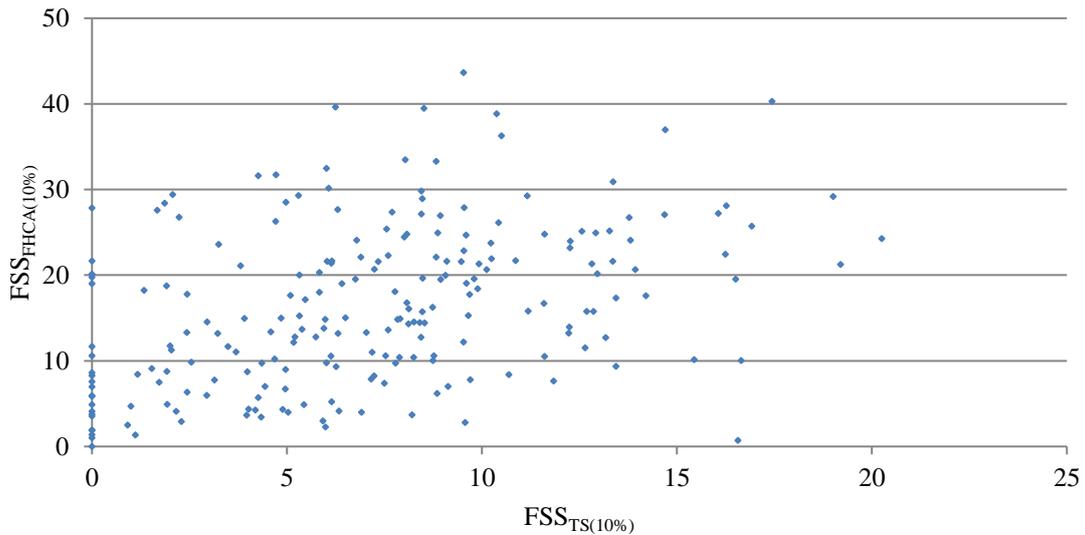

To provide an overall, summary representation of the strengths and weaknesses at field level, in Table 10 we present the union of two sets of SDSs. The first set is composed of the SDSs that place in the second quadrant (high-high) of the distributions for $FSS_{TS(5\%)}$ and $FSS_{FHCA(5\%)}$, taking the borders of the quadrants as the respective medians; the second set is the equivalent for $HCAs_{(10\%)}$. The 58 SDSs identified represent the strongest fields of the Italian research system. Among these we can observe the complete absence of SDSs in Mathematics and the conspicuous presence of SDSs of Chemistry (64%), Biology (42%), Physics (38%), Industrial and information engineering (33%), and Medicine (30%).

With the same procedure, we identify the union of the sets of SDSs that place in the first quadrant (low-low), meaning the 53 weakest fields of our research system (Table 11). Among these, we note the complete absence of SDSs in Chemistry and the conspicuous presence of SDSs of Mathematics (70%), Economics and statistics (65%), and Agricultural and veterinary sciences (30%). Physics and Medicine, in particular, demonstrate an inconsistency in the disciplines, in which fields that excel are flanked by extremely weak ones, 25% and 24% respectively.

Finally, Table 12 provides a summary list of the top 10 (4.6%) and bottom 10 (4.6%) SDSs considering the average of rankings in the four distributions. The table confirms the strengths in some fields of Physics (12.5%), Medicine (8.0%), and Industrial and information engineering (7.1%), to which it adds Psychology and pedagogy (10%). Weaknesses are confirmed in Mathematics (30%) and Economics and statistics (5.9%), while inconsistency applies to Industrial and information engineering (7.1% of SDSs among the weakest 10).

*Table 10: Fields showing all four Fractional Scientific Strength scores above the relevant medians.*

| SDS | UDA* | $FSS_{TS}(5\%)$ | $FSS_{TS}(10\%)$ | $FSS_{FHCA}(5\%)$ | $FSS_{FHCA}(10\%)$ |
|---|---|---|---|---|---|
| FIS/03 - Physics of matter | 2 | 6.50 | 10.69 | 9.54 | 22.85 |
| FIS/05 - Astronomy and Astrophysics | 2 | 8.07 | 20.44 | 10.38 | 38.84 |
| FIS/07 - Applied Physics (Cultural Heritage, Environment, Biology …) | 2 | 11.32 | 9.02 | 8.94 | 19.48 |
| CHIM/01 - Analytical Chemistry | 3 | 6.83 | 8.63 | 10.24 | 21.91 |
| CHIM/02 - Physical Chemistry | 3 | 10.85 | 9.65 | 12.27 | 23.97 |
| CHIM/03 - General and Inorganic Chemistry | 3 | 7.10 | 9.89 | 8.87 | 24.94 |
| CHIM/06 - Organic chemistry | 3 | 6.52 | 6.66 | 8.48 | 19.65 |
| CHIM/08 - Pharmaceutical Chemistry | 3 | 6.04 | 6.95 | 9.92 | 21.33 |
| CHIM/09 - Applied Technological Pharmaceutics | 3 | 6.61 | 10.00 | 7.55 | 25.40 |
| CHIM/12 - Environmental Chemistry and Chem. for Cultural Heritage | 3 | 9.43 | 12.16 | 12.57 | 25.11 |
| GEO/04 - Physical Geography and Geomorphology | 4 | 7.60 | 8.53 | 7.60 | 22.31 |
| GEO/05 - Applied Geology | 4 | 8.96 | 7.00 | 13.44 | 17.33 |
| GEO/12 - Oceanography and Atmospheric Physics | 4 | 9.60 | 7.54 | 19.19 | 21.23 |
| BIO/09 - Physiology | 5 | 5.92 | 7.97 | 10.86 | 21.68 |
| BIO/10 - Biochemistry | 5 | 7.51 | 8.33 | 9.10 | 21.62 |
| BIO/11 - Molecular Biology | 5 | 8.75 | 14.24 | 9.55 | 27.89 |
| BIO/12 - Clinical Biochemistry and Biology | 5 | 10.44 | 7.67 | 16.25 | 22.44 |
| BIO/13 - Applied Biology | 5 | 7.66 | 11.40 | 8.93 | 26.97 |
| BIO/14 - Pharmacology | 5 | 7.27 | 8.47 | 9.59 | 24.67 |
| BIO/17 - Histology | 5 | 9.93 | 15.26 | 8.83 | 33.29 |
| BIO/19 - General Microbiology | 5 | 7.79 | 10.20 | 20.25 | 24.28 |
| MED/01 - Medical Statistics | 6 | 11.94 | 12.48 | 10.23 | 23.74 |
| MED/03 - Medical Genetics | 6 | 14.48 | 11.70 | 8.44 | 27.14 |
| MED/04 - General Pathology | 6 | 11.14 | 14.61 | 13.36 | 30.91 |
| MED/07 - Microbiology and Clinical Microbiology | 6 | 7.77 | 7.35 | 7.77 | 18.09 |
| MED/08 - Pathological Anatomy | 6 | 5.41 | 9.58 | 9.47 | 21.60 |
| MED/09 - Internal Medicine | 6 | 7.40 | 10.20 | 8.83 | 22.11 |
| MED/10 - Respiratory Diseases | 6 | 14.86 | 9.73 | 16.51 | 19.54 |
| MED/11 - Cardiovascular Diseases | 6 | 14.58 | 11.34 | 13.81 | 24.09 |
| MED/12 - Gastroenterology | 6 | 12.63 | 13.24 | 13.78 | 26.72 |
| MED/13 - Endocrinology | 6 | 10.48 | 10.78 | 16.92 | 25.70 |
| MED/15 - Blood Diseases | 6 | 15.50 | 23.19 | 17.44 | 40.28 |
| MED/26 - Neurology | 6 | 10.77 | 10.45 | 8.08 | 24.79 |
| MED/39 - Child Neuropsychiatry | 6 | 8.13 | 13.80 | 16.27 | 28.10 |
| MED/48 - Neuropsychiatric and Rehabilitation Nursing | 6 | 16.90 | 13.11 | 8.45 | 29.82 |
| MED/49 - Applied Dietary Sciences | 6 | 13.03 | 10.66 | 10.43 | 26.12 |
| AGR/04 - Horticulture and Floriculture | 7 | 8.13 | 5.92 | 8.13 | 16.07 |

| SDS | UDA | | | | |
|---|---|---|---|---|---|
| AGR/08 - Agrarian Hydraulics and Hydraulic Forest Management | 7 | 19.01 | 13.98 | 19.01 | 29.17 |
| AGR/16 - Agricultural Microbiology | 7 | 12.90 | 9.55 | 11.61 | 24.77 |
| ICAR/04 - Road, Railway and Airport Construction | 8 | 8.55 | 10.09 | 12.82 | 21.33 |
| ICAR/08 - Construction Science | 8 | 9.04 | 8.91 | 9.60 | 19.06 |
| ING-IND/09 - Energy and Environmental Systems | 9 | 5.72 | 25.09 | 9.53 | 43.66 |
| ING-IND/16 - Production Technologies and Systems | 9 | 12.43 | 15.81 | 14.69 | 27.07 |
| ING-IND/17 - Industrial and Mechanical Plant | 9 | 6.04 | 12.03 | 7.25 | 20.70 |
| ING-IND/22 - Science and Technology of Materials | 9 | 14.53 | 11.94 | 16.06 | 27.19 |
| ING-IND/24 - Principles of Chemical Engineering | 9 | 8.61 | 9.56 | 12.92 | 24.92 |
| ING-IND/27 - Industrial and Technological Chemistry | 9 | 5.88 | 18.82 | 14.70 | 36.98 |
| ING-IND/32 - Electrical Convertors, Machines and Switches | 9 | 8.03 | 16.63 | 8.03 | 33.49 |
| ING-IND/33 - Electrical Energy Systems | 9 | 11.81 | 10.82 | 10.12 | 20.66 |
| ING-IND/34 - Industrial Bioengineering | 9 | 11.16 | 13.09 | 11.16 | 29.26 |
| ING-INF/02 - Electromagnetic Fields | 9 | 8.58 | 6.36 | 12.86 | 15.76 |
| ING-INF/03 - Telecommunications | 9 | 7.47 | 13.06 | 8.01 | 24.46 |
| ING-INF/04 - Systems and control engineering | 9 | 8.47 | 14.97 | 8.47 | 28.92 |
| ING-INF/05 - Data Processing Systems | 9 | 8.81 | 10.80 | 9.07 | 19.98 |
| ING-INF/07 - Electric and Electronic Measurement Systems | 9 | 7.20 | 9.38 | 12.96 | 20.16 |
| M-PSI/02 - Psychobiology and Physiological Psychology | 11 | 10.50 | 16.05 | 10.50 | 36.26 |
| M-PSI/04 - Psychology of Development and Psychology of Education | 11 | 6.73 | 13.21 | 7.69 | 27.35 |
| SECS-P/05 - Econometrics | 13 | 11.65 | 6.71 | 8.74 | 16.26 |

\* 1, Mathematics and computer sciences; 2, Physics; 3, Chemistry; 4, Earth sciences; 5, Biology; 6, Medicine; 7, Agricultural and veterinary sciences; 8, Civil engineering; 9, Industrial and information engineering; 10, Pedagogy and psychology; 11, Economics and statistics

*Table 11: Fields showing all four Fractional Scientific Strength scores below the relevant medians*

| SDS | UDA* | $FSS_{TS}(5\%)$ | $FSS_{TS}(10\%)$ | $FSS_{FHCA}(5\%)$ | $FSS_{FHCA}(10\%)$ |
|---|---|---|---|---|---|
| MAT/01 - Mathematical Logic | 1 | 0.00 | 2.16 | 4.34 | 3.43 |
| MAT/02 - Algebra | 1 | 1.11 | 0.64 | 1.11 | 1.36 |
| MAT/03 - Geometry | 1 | 0.91 | 0.86 | 0.91 | 2.50 |
| MAT/04 - Complementary Mathematics | 1 | 0.00 | 0.00 | 2.30 | 2.91 |
| MAT/05 - Mathematical Analysis | 1 | 1.89 | 1.69 | 4.01 | 4.38 |
| MAT/06 - Probability and Mathematical Statistics | 1 | 0.00 | 1.63 | 5.92 | 3.00 |
| MAT/07 - Mathematical Physics | 1 | 1.22 | 1.68 | 4.26 | 5.71 |
| FIS/06 - Physics for Earth and Atmospheric Sciences | 2 | 0.00 | 5.12 | 5.95 | 13.79 |
| FIS/08 - Didactics and History of Physics | 2 | 0.00 | 2.69 | 0.00 | 4.88 |
| GEO/01 - Palaeontology and Palaeoecology | 4 | 1.92 | 2.34 | 1.92 | 8.77 |
| GEO/02 - Stratigraphic and Sedimentological Geology | 4 | 2.70 | 4.52 | 5.39 | 13.67 |
| GEO/11 - Applied Geophysics | 4 | 0.00 | 2.14 | 0.00 | 6.96 |
| BIO/07 - Ecology | 5 | 3.55 | 5.29 | 5.32 | 15.23 |
| MED/19 - Plastic Surgery | 6 | 0.00 | 1.67 | 2.44 | 6.35 |
| MED/20 - Pediatric and Infant Surgery | 6 | 0.00 | 0.71 | 0.00 | 3.77 |
| MED/27 - Neurosurgery | 6 | 0.00 | 1.78 | 4.97 | 8.98 |
| MED/28 - Odonto-Stomalogical Diseases | 6 | 3.11 | 1.50 | 4.44 | 6.99 |
| MED/29 - Maxillofacial Surgery | 6 | 2.72 | 0.80 | 5.43 | 4.89 |
| MED/30 - Eye Diseases | 6 | 3.50 | 4.32 | 6.13 | 10.55 |
| MED/31 - Otorinolaringology | 6 | 0.00 | 0.86 | 3.96 | 3.65 |
| MED/32 - Audiology | 6 | 3.46 | 1.25 | 6.91 | 3.98 |
| MED/42 - General and Applied Hygiene | 6 | 4.60 | 4.46 | 5.17 | 12.17 |
| MED/43 - Legal Medicine | 6 | 1.63 | 1.61 | 4.89 | 4.33 |
| MED/44 - Occupational Medicine | 6 | 4.31 | 4.90 | 5.74 | 12.78 |
| MED/47 - Nursing and Midwifery | 6 | 0.00 | 4.91 | 0.00 | 8.62 |
| AGR/02 - Agronomy and Herbaceous Cultivation | 7 | 3.44 | 3.62 | 4.58 | 13.39 |

| SDS | UDA | | | | |
|---|---|---|---|---|---|
| AGR/09 - Agricultural Mechanics | 7 | 2.04 | 4.29 | 2.04 | 11.25 |
| AGR/18 - Animal Nutrition and Feeding | 7 | 3.86 | 1.91 | 1.93 | 4.93 |
| AGR/19 - Special Techniques for Zoology | 7 | 2.65 | 2.93 | 3.98 | 8.72 |
| VET/03 - General Pathology and Veterinary Pathological Anatomy | 7 | 0.00 | 1.23 | 6.33 | 4.12 |
| VET/07 - Veterinary Pharmacology and Toxicology | 7 | 0.00 | 1.67 | 4.19 | 4.24 |
| VET/08 - Clinical Veterinary Medicine | 7 | 0.00 | 0.18 | 5.03 | 4.00 |
| VET/09 - Clinical Veterinary Surgery | 7 | 0.00 | 0.54 | 5.99 | 2.27 |
| VET/10 - Clinical Veterinary Obstetrics and Gynaecology | 7 | 0.00 | 0.43 | 0.00 | 1.92 |
| ICAR/01 - Hydraulics | 8 | 0.00 | 5.64 | 3.91 | 14.95 |
| ICAR/07 - Geotechnics | 8 | 1.23 | 3.32 | 3.70 | 11.02 |
| ING-IND/02 - Naval and Marine construction and installation | 9 | 0.00 | 1.71 | 0.00 | 4.10 |
| ING-IND/03 - Flight Mechanics | 9 | 0.00 | 1.77 | 0.00 | 5.87 |
| ING-IND/20 - Nuclear Measurement Tools | 9 | 0.00 | 1.61 | 0.00 | 1.91 |
| ING-IND/28 - Excavation Engineering and Safety | 9 | 0.00 | 3.51 | 0.00 | 5.92 |
| ING-IND/30 - Hydrocarburants and Fluids of the Subsoil | 9 | 0.00 | 0.00 | 0.00 | 0.00 |
| M-EDF/01 - Teaching Methods for Physical Activities | 11 | 4.69 | 5.25 | 4.69 | 10.22 |
| SECS-P/01 - Political Economy | 13 | 2.90 | 3.54 | 3.14 | 7.77 |
| SECS-P/02 - Economic Policy | 13 | 1.53 | 4.40 | 1.53 | 9.11 |
| SECS-P/03 - Finance | 13 | 1.99 | 2.29 | 0.99 | 4.70 |
| SECS-P/04 - History of Economic Thought | 13 | 0.00 | 0.97 | 0.00 | 1.36 |
| SECS-P/07 - Business Administration | 13 | 0.98 | 3.70 | 1.72 | 7.50 |
| SECS-P/10 - Business Organisation | 13 | 0.00 | 5.07 | 3.49 | 11.67 |
| SECS-P/11 - Economics of Financial Intermediaries | 13 | 0.74 | 2.10 | 2.95 | 5.98 |
| SECS-S/01 - Statistics | 13 | 1.35 | 2.38 | 4.96 | 6.70 |
| SECS-S/03 - Statistics for Economics | 13 | 1.17 | 3.37 | 1.17 | 8.43 |
| SECS-S/05 - Social Statistics | 13 | 2.55 | 4.65 | 2.55 | 9.86 |
| SECS-S/06 - Mathematics for economics, act. Studies and finance | 13 | 0.86 | 1.71 | 2.16 | 4.09 |

\* 1, Mathematics and computer sciences; 2, Physics; 3, Chemistry; 4, Earth sciences; 5, Biology; 6, Medicine; 7, Agricultural and veterinary sciences; 8, Civil engineering; 9, Industrial and information engineering; 10, Pedagogy and psychology; 11, Economics and statistics

*Table 12: Top 10 and bottom 10 SDSs by the average rank in the distributions referred to the four Fractional Scientific Strength indicators considered*

| | | $FSS_{TS(5\%)}$ | | $FSS_{FHCA(5\%)}$ | | $FSS_{TS(10\%)}$ | | $FSS_{FHCA(10\%)}$ | | |
|---|---|---|---|---|---|---|---|---|---|---|
| SDS* | UDA | v.a. | rank | v.a. | rank | v.a. | rank | v.a. | rank | Avg rank |
| MED/15 | 6 | 15.50 | 3 | 23.19 | 2 | 17.44 | 4 | 40.28 | 2 | 3 |
| AGR/08 | 7 | 19.01 | 1 | 13.98 | 18 | 19.01 | 3 | 29.17 | 19 | 10 |
| MED/04 | 6 | 11.14 | 19 | 14.61 | 15 | 13.36 | 21 | 30.91 | 13 | 17 |
| ING-IND/16 | 9 | 12.43 | 13 | 15.81 | 11 | 14.69 | 14 | 27.07 | 31 | 17 |
| ING-IND/22 | 9 | 14.53 | 7 | 11.94 | 36 | 16.06 | 11 | 27.19 | 29 | 21 |
| M-PSI/02 | 11 | 10.50 | 23 | 16.05 | 10 | 10.50 | 44 | 36.26 | 7 | 21 |
| MED/12 | 6 | 12.63 | 12 | 13.24 | 26 | 13.78 | 18 | 26.72 | 34 | 22 |
| MED/39 | 6 | 8.13 | 45 | 13.80 | 19 | 16.27 | 9 | 28.10 | 23 | 24 |
| FIS/05 | 2 | 8.07 | 49 | 20.44 | 4 | 10.38 | 46 | 38.84 | 5 | 26 |
| ING-IND/34 | 9 | 11.16 | 18 | 13.09 | 29 | 11.16 | 41 | 29.26 | 18 | 26 |
| … | | | | | | | | | | |
| ING-IND/30 | 9 | 0.00 | 183 | 0.00 | 217 | 0.00 | 196 | 0.00 | 218 | 203 |
| VET/10 | 7 | 0.00 | 183 | 0.43 | 214 | 0.00 | 196 | 1.92 | 212 | 201 |
| SECS-P/04 | 13 | 0.00 | 183 | 0.97 | 206 | 0.00 | 196 | 1.36 | 214 | 200 |
| MAT/02 | 1 | 1.11 | 177 | 0.64 | 212 | 1.11 | 193 | 1.36 | 215 | 199 |
| MED/20 | 6 | 0.00 | 183 | 0.71 | 211 | 0.00 | 196 | 3.77 | 202 | 198 |
| MAT/03 | 1 | 0.91 | 180 | 0.86 | 208 | 0.91 | 195 | 2.50 | 210 | 198 |
| ING-IND/20 | 9 | 0.00 | 183 | 1.61 | 198 | 0.00 | 196 | 1.91 | 213 | 197 |
| MAT/04 | 1 | 0.00 | 183 | 0.00 | 217 | 2.30 | 178 | 2.91 | 208 | 196 |
| ING-IND/02 | 9 | 0.00 | 183 | 1.71 | 190 | 0.00 | 196 | 4.10 | 198 | 192 |

| | | | | | | | | | |
|---|---|---|---|---|---|---|---|---|---|
| MED/31 | 6 | 0.00 | 183 | 0.86 | 209 | 3.96 | 164 | 3.65 | 204 | 190 |

\* MED/15 - Blood Diseases; AGR/08 - Agrarian Hydraulics and Hydraulic Forest Management; MED/04 - General Pathology; ING-IND/16 - Production Technologies and Systems; ING-IND/22 - Science and Technology of Materials; M-PSI/02 - Psychobiology and Physiological Psychology; MED/12 - Gastroenterology; MED/39 - Child Neuropsychiatry; FIS/05 - Astronomy and Astrophysics; ING-IND/34 - Industrial Bioengineering; ING-IND/30 - Hydrocarburants and Fluids of the Subsoil; VET/10 - Clinical Veterinary Obstetrics and Gynaecology; SECS-P/04 - History of Economic Thought; MAT/02 - Algebra; MAT/03 - Geometry; MED/20 - Pediatric and Infant Surgery; ING-IND/20 - Nuclear Measurement Tools; MAT/04 - Complementary Mathematics; ING-IND/02 - Naval and Marine construction and installation; MED/31 – Otorinolaringology.

## 5. Conclusions

The ability to produce significant knowledge advances (HCAs) is a hallmark of scientific excellence. For the same resources employed, the more numerous these advances in a field and the scientists able to produce them, the more that field can be considered a strength of the country research system. The bibliometric indicators proposed in this paper seek to measure these two closely related dimensions of excellence.

The application of the indicators to the Italian academic research system enables identification of strengths and weaknesses at the level of fields and disciplines. The investigation reveals that national strengths in fields of Physics and Medicine are flanked by weaknesses in fields of Mathematics and Economics and statistics; while inconsistency applies to Industrial and information engineering, which presents both strong and weak fields.

The use of results from strategic analyses such as those carried out are twofold. On the one hand they can inform research (and industrial) policy, on the other hand they allow assessment of the effectiveness of any policy initiatives adopted to orient research. The next National Programme for Research 2020-2025, currently under development in Italy, could benefit significantly from the results of the strategic analysis conducted: which of the scientific fields of weakness are considered strategic for the competitiveness of the country system and the socio-economic development in the medium to long term? And which of the fields of strength are not? Is there an alignment between the country's scientific fields of excellence and its industrial sectors of excellence, prefiguring support from the former to the latter, through knowledge transfer and higher education? In short, does the supply of knowledge by universities meets the relevant demand by industry?

Furthermore, what is the world market share (scientific production) of the country in the strong and weak fields? Are these shares growing or decreasing? And are the other countries belonging to the same competitive group investing or divesting in sectors where the country is strong/weak?

The answer to such questions allows the formulation of informed (industrial) research policies and priorities of intervention.

The proposed methodology for identifying the scientific strengths and weaknesses of a country avoids distortions related to the size dependency of traditional methodologies, i.e. those measuring the share of a country's articles, citations, or highly-cited articles relative to the world total. Furthermore, as compared to our previous methodology (Abramo, & D'Angelo, 2014), the methods adopted to control for varying intensity of publication and co-authorships across fields, and differing the threshold used to identify the HCAs, should lead to higher accuracy and substantial robustness. A comparison of results between the previous and current study shows that only three of the top ten subject categories, in terms of scientific strength, remain among the top when applying the improved methodology. It must be said that difference in results might be partly explained by the different periods of observation, 2006-2010 vs 2012-2016. However, for a nation, developing scientific strength in a research field requires a very long time. At the same time, it cannot be expected to wane abruptly in normal conditions. A decisive improvement in the proposed methodology could be realized through international comparisons of research performance at field level. That will very much depend on how rapidly RIMs will be adopted and purposely adapted to the aim, by an ever higher number of countries. In a work in progress aimed at comparing research performance of Italian and Norwegian academics, we have already been able to compare nation-level research performance in each subject category, but more nations should be included in the analysis to come up with a robust benchmark.

Of course, for this and future methods, all the limits and approximations embedded in bibliometric analysis remain, namely: i) publications are not representative of all knowledge produced; ii) bibliometric repertories do not cover all publications; and iii) citations are not always positive, certification of real use, or representative of all use. Finally, results are sensitive to the field classification scheme of researchers, and to the conventions adopted, i.e. the definition of top scientists.